\documentclass[letterpaper]{article} 
\usepackage{aaai24}  
\usepackage{times}  
\usepackage{helvet}  
\usepackage{courier}  
\usepackage[hyphens]{url}  
\usepackage{graphicx} 
\usepackage{natbib}  
\usepackage{caption} 
\usepackage{multirow}
\usepackage{amssymb}
\usepackage{color}
\usepackage{algorithm}
\usepackage{algorithmic}

\usepackage{newfloat}
\usepackage{listings}
\DeclareCaptionStyle{ruled}{labelfont=normalfont,labelsep=colon,strut=off} 
\floatstyle{ruled}
\newfloat{listing}{tb}{lst}{}
\floatname{listing}{Listing}
\title{Multichannel AV-wav2vec2: A Framework for Learning Multichannel Multi-Modal Speech Representation}
\author{
    Qiushi Zhu\textsuperscript{\rm 1}, 
    Jie Zhang\textsuperscript{\rm 1}\thanks{Corresponding author: Jie Zhang.},
    Yu Gu\textsuperscript{\rm 2},
    Yuchen Hu\textsuperscript{\rm 3},
    Lirong Dai\textsuperscript{\rm 1}\\
}
\affiliations{
    \textsuperscript{\rm 1}NERC-SLIP, University of Science and Technology of China (USTC), Hefei, China\\
    \textsuperscript{\rm 2}Tencent AI LAB\\
    \textsuperscript{\rm 3}Nanyang Technological University, Singapore\\
    
}

\usepackage{bibentry}

\begin{document}

\maketitle

\begin{abstract}
Self-supervised speech pre-training methods have developed rapidly in recent years, which show to be very effective for many near-field single-channel speech tasks. However, far-field multichannel speech processing is suffering from the scarcity of labeled multichannel data and complex ambient noises.
The efficacy of self-supervised learning for far-field multichannel and multi-modal speech processing has not been well explored.
Considering that visual information helps to improve speech recognition performance in noisy scenes, in this work we propose a multichannel multi-modal speech self-supervised learning framework AV-wav2vec2, which utilizes video and multichannel audio data as inputs.
First, we propose a multi-path structure to process multichannel audio streams and a visual stream in parallel, with intra- and inter-channel contrastive losses as training targets to fully exploit the spatiotemporal information in multichannel speech data.
Second, based on contrastive learning, we use additional single-channel audio data, which is trained jointly to improve the performance of speech representation.
Finally, we use a Chinese multichannel multi-modal dataset in real scenarios  to validate the effectiveness of the proposed method on audio-visual speech recognition (AVSR), automatic speech recognition (ASR), visual speech recognition (VSR) and audio-visual speaker diarization (AVSD) tasks. 

\end{abstract}

\section{Introduction}

\label{sec:intro}
Over the past few years, speech devices have become ubiquitous. A variety of technologies have emerged to allow them to better recognize speech in noisy environments. According to the distance from the speaker to the device, it can be divided into near-field and far-field automatic speech recognition (ASR)~\cite{li2014overview}. Considering the number of microphones for collecting signals, one can design single-channel ASR~\cite{gulati20_interspeech} and multichannel ASR~\cite{xiao2016deep,ochiai2017multichannel,tu2019iterative,chang2021end,park2020robust} systems. With the recently developed deep learning techniques, near-field single-channel ASR models have been able to achieve satisfactory performance in various environments, while far-field  multichannel ASR still remains challenging, particularly in noisy scenes.

For  multichannel ASR, there are two mainstream approaches: 1) a front-end beamformer in combination with a back-end single-channel ASR model~\cite{xiao2016deep,tu2019iterative,kumatani2019multi,kumatani2012microphone}, and 2) end-to-end multichannel schemes~\cite{ganapathy20183,chang2021end,park2020robust,chang21_interspeech}.
In the former category, most methods are devoted to front-end beamforming, where multichannel recordings are first fed to the beamformer to obtain a single-channel speech, which is then input to the back-end single-channel ASR model.
Multichannel neural filtering methods include fixed beamforming~\cite{kumatani2019multi,minhua2019frequency} and adaptive beamforming~\cite{li2016neural,heymann2016neural}.
The weights of fixed beamformers are fixed in the inference stage, while that of adaptive beamformers can vary with input utterances or noise statistics.
The front-end beamformer and back-end ASR can be trained separately or jointly to improve the  model capacity.
All these methods can improve the multichannel ASR performance to some extent, but the problem is that the training of the front-end beamformer depends on the use of clean speech data and synthesized speech data, and the noise statistics of the synthesized speech may significantly differ from those of the real speech, which might be detrimental to the back-end single-channel ASR.

\begin{figure*}[!ht]
\centering
\includegraphics[width=0.85\textwidth]{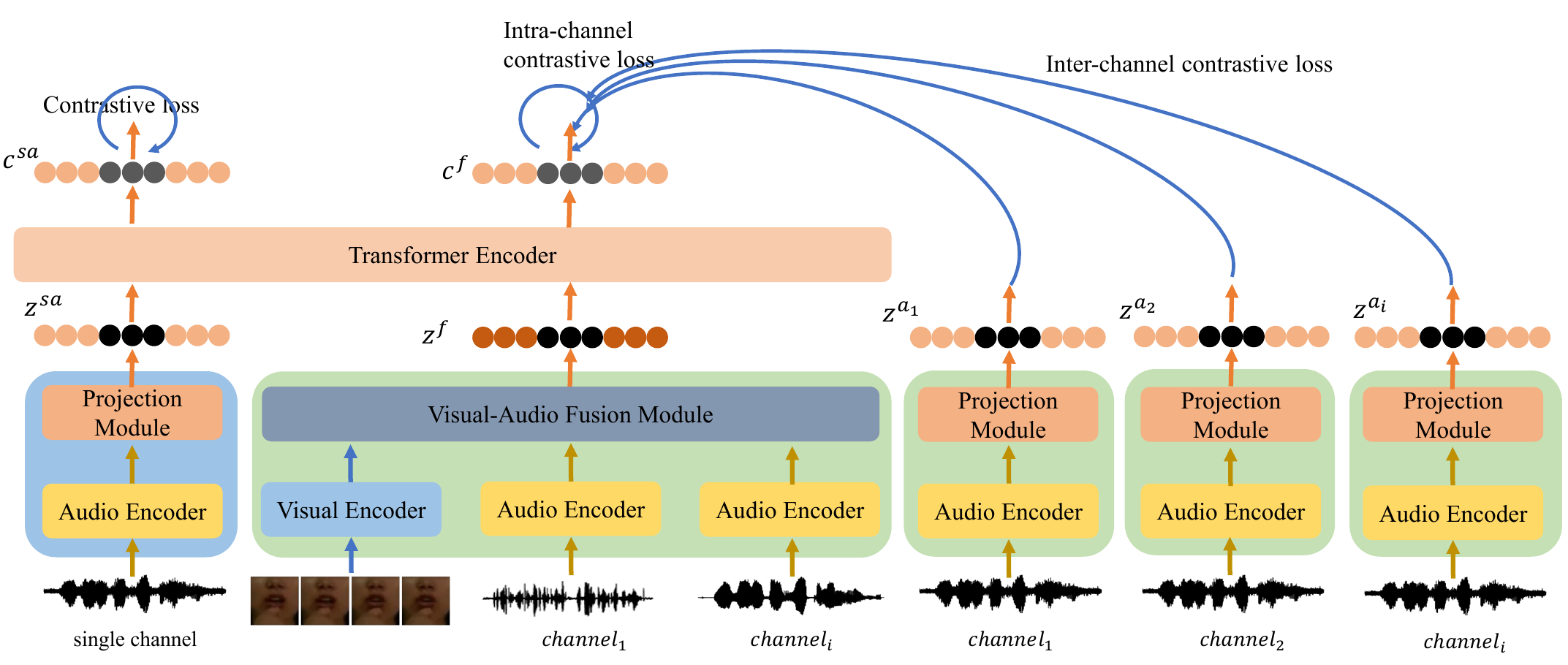}
\caption{The structure of our proposed multichannel multi-modal speech self-supervised pre-training framework.}
\label{fig:figure1}
\end{figure*}

End-to-end multichannel ASR models~\cite{ganapathy20183,chang2021end,park2020robust,chang21_interspeech} are thus proposed to skip the beamforming step and directly utilize the raw multichannel speech data for modeling, since the raw data contains rich real speech and noise statistics.
It was shown in the literature that the end-to-end approach is superior to the front-end and back-end cascade counterparts.
For example, three-dimensional convolutional neural networks (3-D CNNs) have been used for multichannel end-to-end ASR, which outperform beamformer-based and 2-D CNN-based methods~\cite{ganapathy20183}.
Chang et al.~\cite{chang2021end} utilized a multichannel transformer model for ASR, which outperforms the single-channel transformer model as well as the beamformer cascaded with the single-channel transformer model.
Subsequently, the multichannel transformer transducer was proposed in~\cite{chang21_interspeech}, which has a better performance than the multichannel transformer and was shown to be more suitable for streaming ASR.
But these end-to-end multichannel ASR models suffer from the small amount of multichannel speech data as well as little labeled speech data.
As the currently prevalent self-supervised methods can substantially improve the performance of single-channel ASR, {\it how to incorporate self-supervised representation learning in the context  of multichannel ASR remains unexplored}.

There are many popular self-supervised models  proposed for single-channel speech processing, e.g., Wav2vec2.0~\cite{NEURIPS2020_92d1e1eb}, HuBERT~\cite{9585401} as well as their variants like WavLM~\cite{9814838}, SpeechUT~\cite{zhang2022speechut}, etc.
In order to improve the ASR accuracy in noisy conditions, information from additional modalities can be utilized, e.g., video. Recent works~\cite{shi2022robust,9414567,ma2022visual,hu2023cross} show that the visual information can effectively improve the performance, particularly in low signal-to-noise ratio (SNR) scenarios. 
Self-supervised pre-training models based on the visual-audio modality validate the effectiveness of the method on audio-visual speech recognition (AVSR) and visual speech recognition (VSR) tasks~\cite{shi2022learning,zhu2022vatlm,lian2023av,hu2023hearing,hu2023mir}.
These models are able to utilize large amounts of unlabeled unimodal/multi-modal data to substantially improve the ASR performance in noisy scenes.
However, {\it there is a lack of investigating the applicability of these methods  to the multichannel multi-modal context}.

In order to better leverage multichannel or multi-modal multichannel speech information in noisy scenarios, inspired by~\cite{NEURIPS2020_92d1e1eb,shi2022learning,zhu2022vatlm}, we first propose a  multichannel multi-modal self-supervised AV-wav2vec2 model, which is built upon a contrastive learning approach using intra- and inter-channel contrastive loss functions.
Since different speech channels  contain the same speech content but different noise content, we utilize speech signals from different channels to provide self-supervised pre-training targets.
Specifically, the fused video and audio features are fed into the transformer encoder, and then the audio features from each channel provide the self-supervised pre-training target. It is worth mentioning that all audio encoders share the same model parameters.
Second, to tackle the problem of small amounts of multichannel multi-modal data, we utilize additional unlabeled single-channel audio data for multi-task joint training to improve the performance of multi-modal representation learning.
Each batch contains different data types, and the corresponding contrastive loss functions are computed for different data types.
We validate the model performance by fine-tuning the pre-trained model on several downstream tasks, including AVSR, ASR and VSR. Experimental results show consistent performance gains obtained on far-field, mid-field and near-field data.
In addition, the proposed model is further evaluated on the audio-visual speaker diarization (AVSD) task by using the pre-trained model as a feature extractor.
The pre-processed data and code are available at \url{https://github.com/zqs01/multi-channel-wav2vec2}.

\begin{figure}[!t]
	\centering	\includegraphics[width=0.42\textwidth]{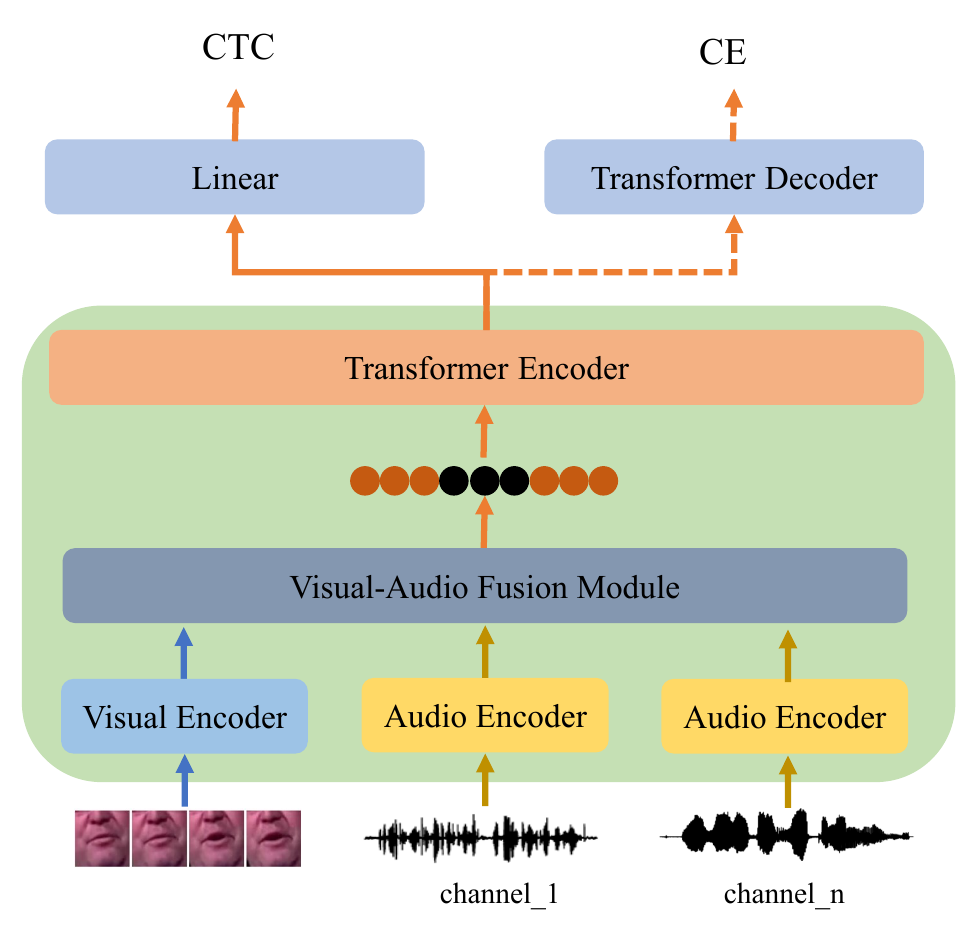}
	\caption{The overall model structure for the downstream AVSR, ASR, and VSR tasks.}
	\label{fig:figure2}
\end{figure}

\section{Methodology}
\subsection{Model Structure}
The proposed multichannel multi-modal self-supervised pre-training framework mainly consists of a multichannel audio-visual branch and a single-channel audio branch, which is shown in Fig.~\ref{fig:figure1}.
The multichannel audio-visual branch  contains a visual encoder, audio encoder, visual-audio fusion module, and Transformer encoder module.
The visual encoder utilizes a modified Resnet-18, where the first convolutional layer is replaced by a 3D convolution with a kernel size of 5$\times$7$\times$7, and a 2D average pooling layer is then used to flatten each video frame into a 1D vector.
Specifically, we extract a 96$\times$96  region-of-interest (ROI) centered by the mouth from the video using Dlib~\cite{king2009dlib} to detect 68 facial key points, which is then fed into the visual encoder to obtain the visual feature $z^v$.
The audio encoder utilizes eight convolutional layers and its output features have a frame shift of 40 ms.
Since the video frame rate is 25 fps, the video and audio features have the same length.
For multichannel audio, we share the audio encoder to provide multichannel targets for representation learning, and let the feature of the $i$-th audio channel be denoted by $z^{a_i}$.
The visual-audio fusion module uses a simple concatenation layer, and the fused features are given by $z^{f}={\rm concat}(z^v,z^{a_1},\cdots,z^{a_i})$.
The masked fusion feature $z^f$ is fed into the Transformer encoder to learn the contextual representation $c^{f}$. 
As it was shown in~\cite{zhu2022vatlm} that utilizing more unlabeled audio data can improve the performance of multi-modal representation, in addition to these multichannel audio-video data, we therefore employ extra single-channel unlabeled audio data. The single-channel audio is fed into the shared audio encoder to obtain the feature $z^{sa}$, and the masked feature $z^{sa}$ is then fed into the shared transformer encoder to obtain the corresponding audio representation $c^{sa}$.

In the pre-training stage, we use data augmentation techniques for the video and audio data.
Similarly to~\cite{shi2022learning}, modality dropout is applied, which performs a random dropout for the visual and audio branches.
We also perform a random dropout for the audio branches of different channels, so as to simulate a realistic scenario of modality missing.
Since we consider six microphone channels for training, the model can be used in any scenario with less than six channels.
In addition, we dynamically add noise to each branch to improve the model's robustness in noisy scenarios.

\subsection{Self-supervised Pre-training Task}
In order to improve the efficiency of pre-training and better leverage the spatial information across different  microphone channels, we use both intra- and inter-channel contrastive loss functions.
In fact, we have considered several mainstream self-supervised pre-training methods, such as the AV-HuBERT and VATLM models that require multiple clustering and multiple iterations, which take a long training time.
Mean-teacher-dependent models such as AV-data2vec and RAVEn only need to be trained once, which however might be unstable and collapse easily during the training process due to the update procedure.
Therefore, we use a self-supervised method based on contrastive learning, which requires training only once and is thus more stable when the model is trained.
As shown in Fig.~\ref{fig:figure1}, at the masking time step $t$, given the predicted output $c^f_t$ and the target $z^f_t$, $z^{a_1}_t,\cdots,z^{a_i}_t$, the intra-channel contrastive loss $\mathcal{L}_{c1}$ and inter-channel contrastive loss  $\mathcal{L}_{c2}$ are respectively given by 
\begin{equation}
\mathcal{L}_{c1}=-\log\frac{\exp(sim(c^f_t, z^f_t)/\kappa)}{\sum_{\tilde{z}^f\sim\{z^f_t,z^f_n\}} \exp(sim(c^f_t,\tilde{z}^f)/\kappa)},
\label{eq1}
\end{equation}
\begin{equation}
 \mathcal{L}_{c2}=-\sum_{i=0}^{5}\log\frac{\exp(sim(c^{f}_t, z^{a_i}_t)/\kappa)}{\sum_{\tilde{z}^{a_i}\sim\{z^{a_i}_t,z^{a_i}_n\}} \exp(sim(c^{f}_t,\tilde{z}^{a_i})/\kappa)},
\label{eq2}
\end{equation}
where $sim$ denotes the cosine similarity, $\kappa$ is the temperature coefficient, $z^f$ denotes the fusion representation, $z^{a_i}$ denotes the feature corresponding to the $i$-th audio channel,
$\tilde{z}^f$ and $\tilde{z}^{a_i}$ are the negative samples of the corresponding branches, $c^f_n$ is the negative sample randomly sampled from the Transformer output representation, and $z^{a_i}_n$ is the negative sample randomly sampled from the output feature of the audio encoder.
The loss function $\mathcal{L}_{sa}$ corresponding to the single-channel audio is given by
\begin{equation}
		\mathcal{L}_{sa}= -\log\frac{\exp(sim(c^{sa}_t, z^{sa_i}_t)/\kappa)}{\sum_{\tilde{c}^{sa_i}\sim\{z^{sa_i}_t,z^{sa_i}_n\}} \exp(sim(c^{sa_i}_t,\tilde{z}^{sa_i})/\kappa)},
\label{eq3}
\end{equation}
where $z^{sa_i}_n$ is the negative samples derived from the output features of the audio encoder.
Therefore, the total loss function $L_{total}$ is given by 
\begin{equation}
\mathcal{L}_{total}=\mathcal{L}_{c1} + \mathcal{L}_{c2} +  \lambda \mathcal{L}_{sa},  \label{eq4}
\end{equation}
where $\lambda$ is the hyperparameter.

\subsection{Downstream Tasks}

As shown in Fig~\ref{fig:figure2}, we perform AVSR, ASR, and VSR tasks by fine-tuning the pre-trained model, as well as AVSD by using the pre-trained model as a feature extractor.
For AVSR, ASR, and VSR tasks, we add Transformer decoder layers or linear layers on top of the pre-trained model and optimize the model using the cross-entropy (CE)~\cite{7472618} or connectionist temporal classification (CTC)~\cite{graves2006connectionist} loss function.
In case one modality is missing, we replace it with a zero vector.

\begin{figure}[!t]
	\centering
	\includegraphics[width=0.45\textwidth]{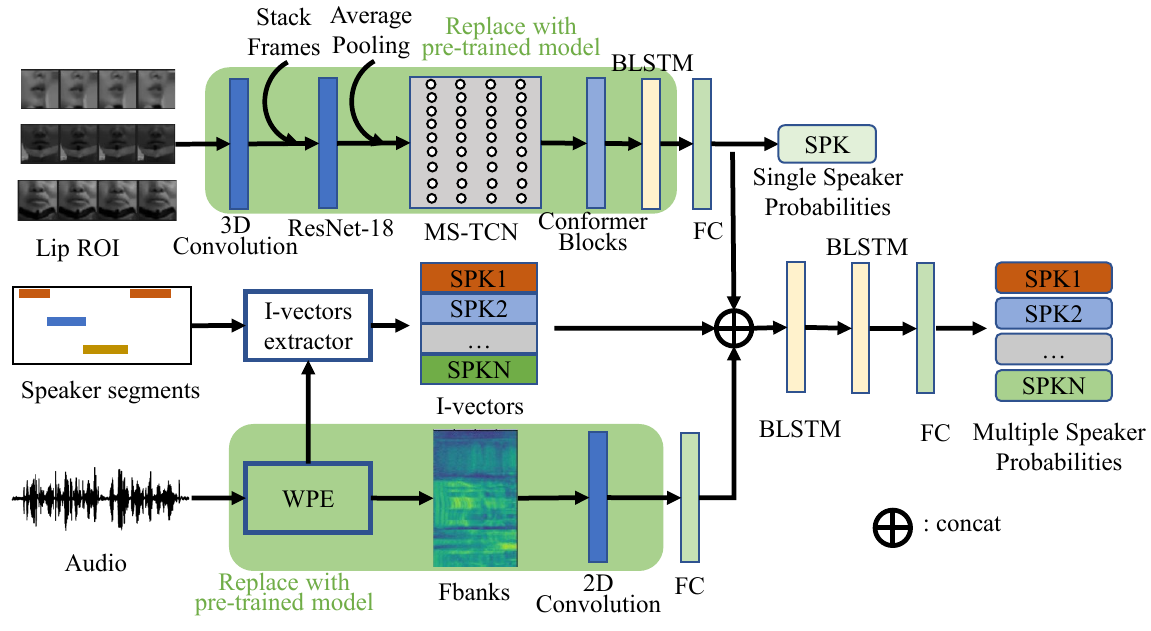}
	\caption{The overall model structure for the downstream AVSD task.}
	\label{fig:figure3}
\end{figure}

For the AVSD task, we use the model structure from~\cite{he22c_interspeech} as the baseline, which is shown in Figure~\ref{fig:figure3}.
The model concatenates the visual embedding branch, audio embedding branch, and speaker embedding branch, and then uses bidirectional long short-term memory (BLSTM)~\cite{graves2012long} network for end-to-end optimization.
For the visual embedding branch, the ROI of multiple speakers is passed through the 3D convolution module, ResNet-18, the multi-scale temporal Convolution (MS-TCN) layer~\cite{9053841}, the conformer~\cite{gulati20_interspeech} layer and the BLSTM layer, and finally the fully connected layer (FC) layer in a step-by-step fashion to obtain the probabilities of individual speakers.
For the speaker embedding branch, the corresponding I-vector features are extracted using the audio of different speakers.
For the audio embedding branch, the multichannel audio is fed into the weighted prediction error (WPE) dereverberation  and then the log-filterbank features are extracted, which are then fed into the 2D convolutional layer and FC layer.
The three branches are concatenated and  fed into the BLSTM and FC layers to obtain the probability of multiple speakers. More details on this baseline model can be found in~\cite{he22c_interspeech}.
We replace the original visual embedding branch or audio embedding branch or both with the pre-trained model with frozen parameters (e.g., see the green part of Fig.~\ref{fig:figure3}).
The hyperparameters and training methods of the replaced model are the same as those of the baseline model.
In addition, we will compare the performance of different pre-trained models (e.g., AV-HuBERT, VATLM) on this task in experiments.

\section{Experimental Setup}

\subsection{Data Description}
\textbf{Pre-training data:}
For the multichannel audio-visual dataset, we utilize the MISP2021-AVSR\footnote{\url{https://challenge.xfyun.cn/misp\_dataset}}~\cite{chen22o_interspeech,9746683} dataset, which contains a total of 141.24 hours of audio and video data collected from 253 Chinese (98 males and 165 females) speaking Mandarin in 34 real home TV rooms.
There is no overlap of speakers and recording rooms for each subset.
Multiple microphones and cameras are placed in the room to collect far/middle/near audio data and far/middle video data simultaneously.
For the audio data, the far recording device consists of a linear microphone array composed of 6 synchronized omnidirectional microphones, which are placed 3-5 meters away from the speaker.
The middle recording device is placed 1-1.5 meters away from the speaker and consists of a 2-sample synchronized omnidirectional microphone.
Each speaker wears a high-fidelity microphone in the middle of the chin to collect near data.
For the video data, a wide-angle camera is placed 3-5 meters away from the speaker to collect far video.
High-definition cameras are placed 1-1.5 meters away from the speaker to collect the middle video.
The recording environment can be controlled during data collection, such as daytime or night, and the TV/lights can be turned on or off.
According to different configurations, the data are divided into four types, which are summarized in Table~\ref{tab:table1}.
For more details on the MISP2021-AVSR dataset, please refer to~\cite{chen22o_interspeech,9746683}.
In addition, we also utilize the single-channel Chinese Mandarin dataset WenetSpeech\footnote{\url{https://github.com/wenet-e2e/WenetSpeech}}~\cite{9746682} in the pre-training phase, which contains 10000+ hours of multi-domain Chinese data.
It is worth noting that we only use the unlabeled audio data from WenetSpeech for pre-training.
We select approximately 900 hours of data from Youtube.

\begin{table}[]
\centering
\setlength{\tabcolsep}{0.45mm}{
\begin{tabular}{l|cccccc|cccccc}
		\hline
		\textbf{ConfigID} & \multicolumn{3}{c|}{\textbf{C1}}                                                         & \multicolumn{3}{c|}{\textbf{C2}}                                    & \multicolumn{3}{c|}{\textbf{C3}}                                                         & \multicolumn{3}{c}{\textbf{C4}}                                     \\ \hline
		Group    & \multicolumn{6}{c|}{1}                                                                                                                       & \multicolumn{6}{c}{2}                                                                                                                        \\ \hline
		TV       & \multicolumn{3}{c|}{off}                                                        & \multicolumn{3}{c|}{on}                                    & \multicolumn{3}{c|}{off}                                                        & \multicolumn{3}{c}{on}                                     \\ \hline
		Time     & \multicolumn{2}{c|}{day}                           & \multicolumn{1}{c|}{night} & \multicolumn{2}{c|}{day}                           & night & \multicolumn{2}{c|}{day}                           & \multicolumn{1}{c|}{night} & \multicolumn{2}{c|}{day}                           & night \\ \hline
		Light    & \multicolumn{1}{c|}{on} & \multicolumn{1}{c|}{off} & \multicolumn{1}{c|}{on}    & \multicolumn{1}{c|}{on} & \multicolumn{1}{c|}{off} & on    & \multicolumn{1}{c|}{on} & \multicolumn{1}{c|}{off} & \multicolumn{1}{c|}{on}    & \multicolumn{1}{c|}{on} & \multicolumn{1}{c|}{off} & on    \\ \hline
	\end{tabular}}
\caption{The summary of the four different types of 
 data subsets based on the recording environment.}
 \label{tab:table1}
\end{table}

\subsection{Data Pre-processing}
For the video data, we extract a 96$\times$96 ROI centered by the mouth from the video using Dlib to detect 68 facial key points, which is then fed into the visual encoder.
For audio data, we convert the audio sample rate to 16 kHz in prior to being fed into the audio encoder.
For the beamforming-based approach, we use WPE dereverberation~\cite{8578026} and weighted delay-and-sum  beamforming (BeamformIt)~\cite{4291588} for the far-field 6-channel speech and the middle-field 2-channel speech before being sent to the model.
Note that the near-field single-channel data keeps in the original form.

\subsection{Fine-tuning and Evaluation}
After pre-training, we fine-tune the pre-trained model or use the pre-trained model as a feature extractor. 
For the AVSR/ASR/VSR task, we add a 6-layer Transformer decoder on top of the pre-trained model and then fine-tune the whole model using labeled data.
The embedding layer dimension of the decoder is set to 768, the layer  dimension of the feed-forward neural network is set to 3072, and the number of attention heads is set to 12.
The Adam optimizer is still utilized and the learning rate is set to 0.001.
We use the label smoothing cross-entropy loss function, where the label smoothing factor is set to 0.1.
The modeling units include about 4000 Chinese characters.
After fine-tuning, we do not use any language model but only use a beam size of 50 to decode the test set.
The character error rate (CER) is used as the evaluation metric.
For the AVSD task, the hyperparameters and training methods of the model are the same as those of the baseline model, and the parameters of the pre-trained model are not updated.
The Adam optimizer is used and the learning rate is set to 1e-4. The objective of the optimization is a binary CE loss function. The diarization error rate (DER) is utilized as the evaluation metric for AVSD.
The baseline code used for the AVSD task is available online\footnote{\url{https://github.com/mispchallenge/misp2022\_baseline/tree/main/track1\_AVSD}}.

\begin{table*}[!ht]
\centering
\begin{tabular}{lc|cc|cccc}
\hline
\multicolumn{2}{l|}{\multirow{2}{*}{\textbf{Method}}}                                                                                                                            & \multicolumn{1}{c|}{\multirow{2}{*}{\textbf{Audio}}} & \multirow{2}{*}{\textbf{Video}} & \multicolumn{4}{c}{\textbf{CER (\%)}}                             \\ \cline{5-8} 
\multicolumn{2}{l|}{}                                                                                                                                                            & \multicolumn{1}{c|}{}                                &                                 & \textbf{C1}    & \textbf{C2}    & \textbf{C3}    & \textbf{C4}    \\ \hline
\multicolumn{1}{l|}{\multirow{25}{*}{Beamforming}}  & \multirow{8}{*}{MS-TCN~\cite{chen22o_interspeech}}                                                                                                    & far                                                  & /                               & 55.47          & 74.70          & 82.19          & 84.21          \\
\multicolumn{1}{l|}{}                               &                                                                                                                            & middle                                               & /                               & 50.51          & 61.79          & 75.05          & 75.83          \\
\multicolumn{1}{l|}{}                               &                                                                                                                            & near                                                 & /                               & 25.71          & 23.45          & 22.59          & 20.89          \\ \cline{3-8} 
\multicolumn{1}{l|}{}                               &                                                                                                                            & /                                                    & far                             & 95.69          & 95.62          & 95.98          & 95.69          \\
\multicolumn{1}{l|}{}                               &                                                                                                                            & /                                                    & middle                          & 89.77          & 88.70          & 89.30          & 87.77          \\ \cline{3-8} 
\multicolumn{1}{l|}{}                               &                                                                                                                            & far                                                  & far                             & 50.99          & 66.21          & 71.55          & 73.73          \\
\multicolumn{1}{l|}{}                               &                                                                                                                            & far                                                  & middle                          & 45.57          & 58.59          & 63.16          & 64.61          \\
\multicolumn{1}{l|}{}                               &                                                                                                                            & near                                                 & middle                          & 22.64          & 20.44          & 19.97          & 18.58          \\ \cline{2-8} 
\multicolumn{1}{l|}{}                               & \begin{tabular}[c]{@{}c@{}}AV-HuBERT-ED~\cite{shi2022learning}\\ (iter5)\end{tabular}                                                             & near                                                 & /                               & x              & x              & x              & x              \\ \cline{2-8} 
\multicolumn{1}{l|}{}                               & \multirow{8}{*}{\begin{tabular}[c]{@{}c@{}}AV-HuBERT-CTC~\cite{shi2022learning}\\ (iter5)\end{tabular}}                                           & far                                                  & /                               & 62.86          & 79.99          & 82.33          & 85.72          \\
\multicolumn{1}{l|}{}                               &                                                                                                                            & middle                                               & /                               & 64.85          & 74.73          & 78.09          & 80.75          \\
\multicolumn{1}{l|}{}                               &                                                                                                                            & near                                                 & /                               & 47.17          & 46.06          & 45.84          & 44.34          \\ \cline{3-8} 
\multicolumn{1}{l|}{}                               &                                                                                                                            & /                                                    & far                             & 90.73          & 91.01          & 90.94          & 89.81          \\
\multicolumn{1}{l|}{}                               &                                                                                                                            & /                                                    & middle                          & 89.91          & 89.85          & 90.11          & 88.85          \\ \cline{3-8} 
\multicolumn{1}{l|}{}                               &                                                                                                                            & far                                                  & far                             & 64.66          & 84.18          & 85.58          & 89.84          \\
\multicolumn{1}{l|}{}                               &                                                                                                                            & far                                                  & middle                          & 66.75          & 85.95          & 86.53          & 91.05          \\
\multicolumn{1}{l|}{}                               &                                                                                                                            & near                                                 & middle                          & 47.87          & 46.82          & 46.44          & 45.05          \\ \cline{2-8} 
\multicolumn{1}{l|}{}                               & \multirow{8}{*}{\begin{tabular}[c]{@{}c@{}}Our AV-wav2vec2-CTC\\ (pre-train 200k)\end{tabular}}                             & far                                                  & /                               & \textbf{44.99} & \textbf{71.03} & \textbf{76.49} & \textbf{81.03} \\
\multicolumn{1}{l|}{}                               &                                                                                                                            & middle                                               & /                               & \textbf{44.48} & \textbf{60.41} & \textbf{66.56} & \textbf{70.74} \\
\multicolumn{1}{l|}{}                               &                                                                                                                            & near                                                 & /                               & \textbf{17.26} & \textbf{16.28} & \textbf{14.93} & \textbf{13.71} \\ \cline{3-8} 
\multicolumn{1}{l|}{}                               &                                                                                                                            & /                                                    & far                             & \textbf{88.99} & \textbf{89.03} & \textbf{88.81} & \textbf{87.99} \\
\multicolumn{1}{l|}{}                               &                                                                                                                            & /                                                    & middle                          & \textbf{86.05} & \textbf{86.73} & \textbf{86.36} & \textbf{85.00} \\ \cline{3-8} 
\multicolumn{1}{l|}{}                               &                                                                                                                            & far                                                  & far                             & \textbf{43.56} & \textbf{65.06} & \textbf{69.65} & \textbf{71.03} \\
\multicolumn{1}{l|}{}                               &                                                                                                                            & far                                                  & middle                          & \textbf{40.34} & \textbf{61.17} & \textbf{66.03} & \textbf{65.90} \\
\multicolumn{1}{l|}{}                               &                                                                                                                            & near                                                 & middle                          & \textbf{17.35} & \textbf{16.11} & \textbf{15.19} & \textbf{14.06} \\ \hline
\multicolumn{1}{l|}{\multirow{16}{*}{Multichannel}} & \multirow{8}{*}{\begin{tabular}[c]{@{}c@{}}Our AV-wav2vec2-CTC\\ (pre-train 200k)\end{tabular}}                             & far                                                  & /                               & 43.74          & 70.38          & 75.6           & 79.57          \\
\multicolumn{1}{l|}{}                               &                                                                                                                            & middle                                               & /                               & 43.56          & 60.09          & 65.94          & 69.02          \\
\multicolumn{1}{l|}{}                               &                                                                                                                            & near                                                 & /                               & 17.03          & 16.15          & 14.88          & 13.56          \\ \cline{3-8} 
\multicolumn{1}{l|}{}                               &                                                                                                                            & /                                                    & far                             & 87.66          & 88.05          & 87.76          & 86.84          \\
\multicolumn{1}{l|}{}                               &                                                                                                                            & /                                                    & middle                          & 85.74          & 86.27          & 85.91          & 84.37          \\ \cline{3-8} 
\multicolumn{1}{l|}{}                               &                                                                                                                            & far                                                  & far                             & 43.01          & 64.78          & 68.94          & 70.65          \\
\multicolumn{1}{l|}{}                               &                                                                                                                            & far                                                  & middle                          & 39.95          & 61.04          & 65.82          & 65.43          \\
\multicolumn{1}{l|}{}                               &                                                                                                                            & near                                                 & middle                          & 17.00          & 16.03          & 15.10          & 14.01          \\ \cline{2-8} 
\multicolumn{1}{l|}{}                               & \multirow{8}{*}{\begin{tabular}[c]{@{}c@{}}Our AV-wav2vec2-CTC\\ (pre-train 200k \\ + 900 hours unlabel data)\end{tabular}} & far                                                  & /                               & \textbf{41.91} & \textbf{68.24} & \textbf{73.27} & \textbf{77.82} \\
\multicolumn{1}{l|}{}                               &                                                                                                                            & middle                                               & /                               & \textbf{42.38} & \textbf{59.13} & \textbf{64.89} & \textbf{68.27} \\
\multicolumn{1}{l|}{}                               &                                                                                                                            & near                                                 & /                               & \textbf{16.57} & \textbf{16.09} & \textbf{14.56} & \textbf{13.45} \\ \cline{3-8} 
\multicolumn{1}{l|}{}                               &                                                                                                                            & /                                                    & far                             & \textbf{86.54} & \textbf{86.73} & \textbf{86.42} & \textbf{84.21} \\
\multicolumn{1}{l|}{}                               &                                                                                                                            & /                                                    & middle                          & \textbf{84.82} & \textbf{85.36} & \textbf{84.75} & \textbf{82.59} \\ \cline{3-8} 
\multicolumn{1}{l|}{}                               &                                                                                                                            & far                                                  & far                             & \textbf{42.33} & \textbf{64.02} & \textbf{67.77} & \textbf{69.47} \\
\multicolumn{1}{l|}{}                               &                                                                                                                            & far                                                  & middle                          & \textbf{39.47} & \textbf{60.45} & \textbf{64.98} & \textbf{64.64} \\
\multicolumn{1}{l|}{}                               &                                                                                                                            & near                                                 & middle                          & \textbf{16.48} & \textbf{15.96} & \textbf{15.02} & \textbf{13.95} \\ \hline
\end{tabular}
\caption{The performance comparison of different methods on AVSR, ASR, and VSR tasks.}
\label{tab:table2}
\end{table*}

\section{Experimental Results}
\subsection{Results on the AVSR/ASR/VSR Task}
\textbf{Comparison method}: Since this Chinese dataset is relatively new, we adopt the method in~\cite{chen22o_interspeech} as the comparison method, which utilizes a MS-TCN  as the backbone for multiple tasks.

The experimental results of different methods on AVSR, ASR and VSR tasks are shown in Table~\ref{tab:table2}.
For the beamforming-based methods, we first merge the multichannel data into single-channel data through the BeamformIt\footnote{\url{https://github.com/xanguera/BeamformIt}} tool, which is then used for model training.
As it was shown that AV-HuBERT (in English) is an excellent audio-visual self-supervised pre-training model, we extend it to the AVSR task for Chinese scenarios.
We pre-train the Chinese AV-HuBERT model using only the  MISP2021-AVSR dataset and strictly follow the steps for training the original AV-HuBERT.
Fine-tuning AV-HuBERT(iter5) does not obtain good results when we add 6 decoder layers to the pre-trained model. We found that in case the amount of Chinese data is small and using 4000 Chinese modeling units is not beneficial for model convergence, as the character error rate (CER) is close to 100\%.
Therefore, we add a linear layer to the pre-trained model and fine-tune the AV-HuBERT(iter5) with the CTC loss function. From Table~\ref{tab:table2}, we find that compared with MS-TCN, the AV-HuBERT-CTC model has some advantages on the far ASR task, but is not good enough for the middle and near cases.
This is due to the fact that in the Chinese context, the large number of modeling units (4000) and the insufficient number of clustering categories may affect the quality of the pre-trained representations.

Considering that pre-training AV-HuBERT for five iterations is time-consuming and resource-consuming, we further propose the AV-wav2vec2 based on contrastive learning, which requires pre-training only once.
In case the CTC loss function is used for fine-tuning AV-wav2vec2, it is clear that the performance is substantially improved on AVSR, ASR, and VSR tasks, where all outperform the  AV-HuBERT-CTC.
In addition, we directly conduct AVSR, ASR, and VSR evaluations using an end-to-end multichannel approach.
In the case of training with random initialization, AV-wav2vec2-CTC (multichannel) performs better than AV-wav2vec2-CTC (beamforming) on the ASR task but performs comparably on the VSR and AVSR tasks.
In the case of joint training with additional 900 hours of unimodal single-channel speech data, the model performance  in far-field scenarios can be further improved, indicating that increasing the amount of unlabeled data is more effective than changing the model itself when the amount of data is limited.
It is interesting that utilizing additional data does not improve the overall performance too much, which is mainly due to the fact that the MISP2021-AVSR dataset was mainly recorded in the local dialect. Including more Chinese data from other areas in this dataset might bring further performance gain.
Due to the page limitation, some experimental results are presented in the Appendix.

\subsection{Evaluation of Intra- and Inter-Channel Contrastive Loss Functions}
In order to verify the effect of different contrast loss functions on the performance, we only utilize the MISP2021-AVSR dataset for pre-training and conduct comparisons on the far-field AVSR task. The results are shown in Table~\ref{tab:table3}.
In case both the intra-channel contrastive loss $\mathcal{L}_{c1}$ and the inter-channel contrastive loss function $\mathcal{L}_{c2}$ are employed, the obtained CER is 70.65 on the C4 test set. 
When only the intra-channel contrastive loss function $\mathcal{L}_{c1}$ is taken into account, the CER becomes 73.12 on the C4 test set. This shows that using features from different channels to provide targets for the self-supervised pre-training model is beneficial.

\subsection{Evaluation on the AVSD Task}
Experimental results of using the pre-trained model as the feature extractor on the AVSD task are shown in Table~\ref{tab:table5}.
In~\cite{wang2023multimodal}, the speaker diarization system using only audio data achieves a DER of 31.25 on the validation set and a DER of 18.69 on the validation set when using only the visual data.
In the case of using both audio and visual data, it achieves a DER of 13.09 on the validation set and 13.88 on the test set, indicating that the bi-modal data can significantly improve the performance compared to the single modality.
We re-train the baseline model incorporating data augmentation and dropout strategies, resulting in a DER of 13.58 on the test set, which is used as our baseline for comparison.

To verify whether pre-trained models can improve the AVSD performance, we first use the publicly available AV-HuBERT model and the VATLM model as feature extractors to replace the audio or visual branch in the baseline model.
When the visual and audio branches of the baseline model are replaced by the AV-HuBERT model, the DERs on the test set decrease to 12.91 and 12.47, respectively. In case the visual and audio branches of the baseline model are both replaced, the DER on the test set reaches 11.96.
We can see that in case the visual and audio branches of the baseline model are replaced by the VATLM model, the corresponding DER is even lower. 
Because AV-HuBERT and VATLM models are trained on English audio-visual datasets, they may suffer from the issue of domain mismatch when extracting features on Chinese audio-visual datasets, but it is interesting to note that these models still bring some performance gains.
Since VATLM utilizes more unlabeled audio and unlabeled text data, it outperforms the AV-HuBERT model.
More importantly, it is clear that if the beamforming-based AV-wav2vec2 model is used to replace the visual and audio branches of the baseline model (either separately or simultaneously), the DER can be further improved compared to the AV-HuBERT case.
This validates the potential and excellent performance of the proposed pre-trained model on the AVSD task.

\begin{table}[!ht]
\centering
\scalebox{0.9}{
\begin{tabular}{l|cc|cccc}
\hline
\multirow{2}{*}{\textbf{Method}} & \multirow{2}{*}{\textbf{Audio}} & \multirow{2}{*}{\textbf{Video}} & \multicolumn{4}{c}{\textbf{CER (\%)}}                 \\
                                 &                                 &                                 & \textbf{C1} & \textbf{C2} & \textbf{C3} & \textbf{C4} \\ \hline
$\mathcal{L}_{c1}+\mathcal{L}_{c2}$                          & far                             & far                             & 43.01       & 64.78       & 68.94       & 70.65       \\
$\mathcal{L}_{c1}$                             & far                             & far                             & 44.68       & 66.94       & 70.98       & 73.12       \\ \hline
\end{tabular}}
\caption{The performance comparison using intra-channel and inter-channel contrast loss functions.}
\label{tab:table3}
\end{table}

\begin{table*}[!ht]
\centering
\scalebox{0.89}{
\begin{tabular}{lcccccccc}
\hline
\multicolumn{1}{l|}{\multirow{2}{*}{\textbf{Method}}}             & \multicolumn{4}{c|}{\textbf{Dev}}                                                 & \multicolumn{4}{c}{\textbf{Eval}}                            \\ \cline{2-9} 
\multicolumn{1}{l|}{}                                             & \textbf{FA} & \textbf{MISS} & \textbf{SPKEER} & \multicolumn{1}{c|}{\textbf{DER}} & \textbf{FA} & \textbf{MISS} & \textbf{SPKEER} & \textbf{DER} \\ \hline
\multicolumn{9}{l}{\textbf{Supervised}}                                                                                                                                                                              \\ \hline
\multicolumn{1}{l|}{ASD~\cite{wang2023multimodal}}                                          & 0.01        & 19.88         & 11.36           & \multicolumn{1}{c|}{31.25}        & -           & -             & -               & -            \\
\multicolumn{1}{l|}{VSD~\cite{wang2023multimodal}}                                          & 6.64        & 8.17          & 3.89            & \multicolumn{1}{c|}{18.69}        & -           & -             & -               & -            \\
\multicolumn{1}{l|}{AVSD~\cite{wang2023multimodal}}                                         & 4.01        & 5.86          & 3.22            & \multicolumn{1}{c|}{13.09}        & -           & -             & -               & 13.88        \\ \hline
\multicolumn{1}{l|}{AVSD (Our baseline)}                          & 3.29        & 6.17          & 3.51            & \multicolumn{1}{c|}{12.97}        & 2.83        & 6.28          & 4.47            & 13.58        \\ \hline
\multicolumn{9}{l}{\textbf{Self-supervised}}                                                                                                                                                                         \\ \hline
\multicolumn{1}{l|}{AVSD (AV-HuBERT visual)}                      & 1.74        & 7.50          & 3.38            & \multicolumn{1}{c|}{12.61}        & 1.99        & 7.06          & 3.86            & 12.91        \\
\multicolumn{1}{l|}{AVSD (AV-HuBERT audio)}                       & 2.07        & 7.03          & 3.24            & \multicolumn{1}{c|}{12.34}        & 2.21        & 6.93          & 3.33            & 12.47        \\
\multicolumn{1}{l|}{AVSD (AV-HuBERT visual + audio)}              & 3.21        & 5.89          & 2.94            & \multicolumn{1}{c|}{12.04}        & 3.04        & 5.64          & 3.28            & 11.96        \\ \hline
\multicolumn{1}{l|}{AVSD (VATLM visual)}                          & 4.19        & 5.18          & 2.81            & \multicolumn{1}{c|}{12.17}        & 3.87        & 5.71          & 2.67            & 12.25        \\
\multicolumn{1}{l|}{AVSD (VATLM audio)}                           & 2.26        & 6.73          & 2.58            & \multicolumn{1}{c|}{11.56}        & 2.46        & 6.69          & 2.65            & 11.80        \\
\multicolumn{1}{l|}{AVSD (VATLM visual + audio)}                  & 1.94        & 6.55          & 2.37            & \multicolumn{1}{c|}{10.85}        & 2.23        & 6.46          & 2.54            & 11.23        \\ \hline
\multicolumn{1}{l|}{AVSD (Beamforming AV-wav2vec2 visual)}         & 3.92        & 4.72          & 2.75            & \multicolumn{1}{c|}{11.39}        & 3.58        & 5.25          & 2.59            & 11.42        \\
\multicolumn{1}{l|}{AVSD (Beamforming AV-wav2vec2 audio)}          & 2.17        & 6.36          & 2.45            & \multicolumn{1}{c|}{10.98}        & 2.39        & 6.37          & 2.60            & 11.36        \\
\multicolumn{1}{l|}{AVSD (Beamforming AV-wav2vec2 visual + audio)} & 1.90        & 6.23          & 2.31            & \multicolumn{1}{c|}{10.44}        & 2.20        & 6.04          & 2.58            & 10.82        \\ \hline
\end{tabular}}
\caption{The performance comparison of supervised and self-supervised methods on the speaker diarization task, where FA, MISS, and SPKEER represent false alarm error, missed detection error, and speaker error, respectively.}
\label{tab:table5}
\vspace{-0.3cm}
\end{table*}

\begin{figure*}[!ht]
\centering	
\includegraphics[width=0.85\textwidth]{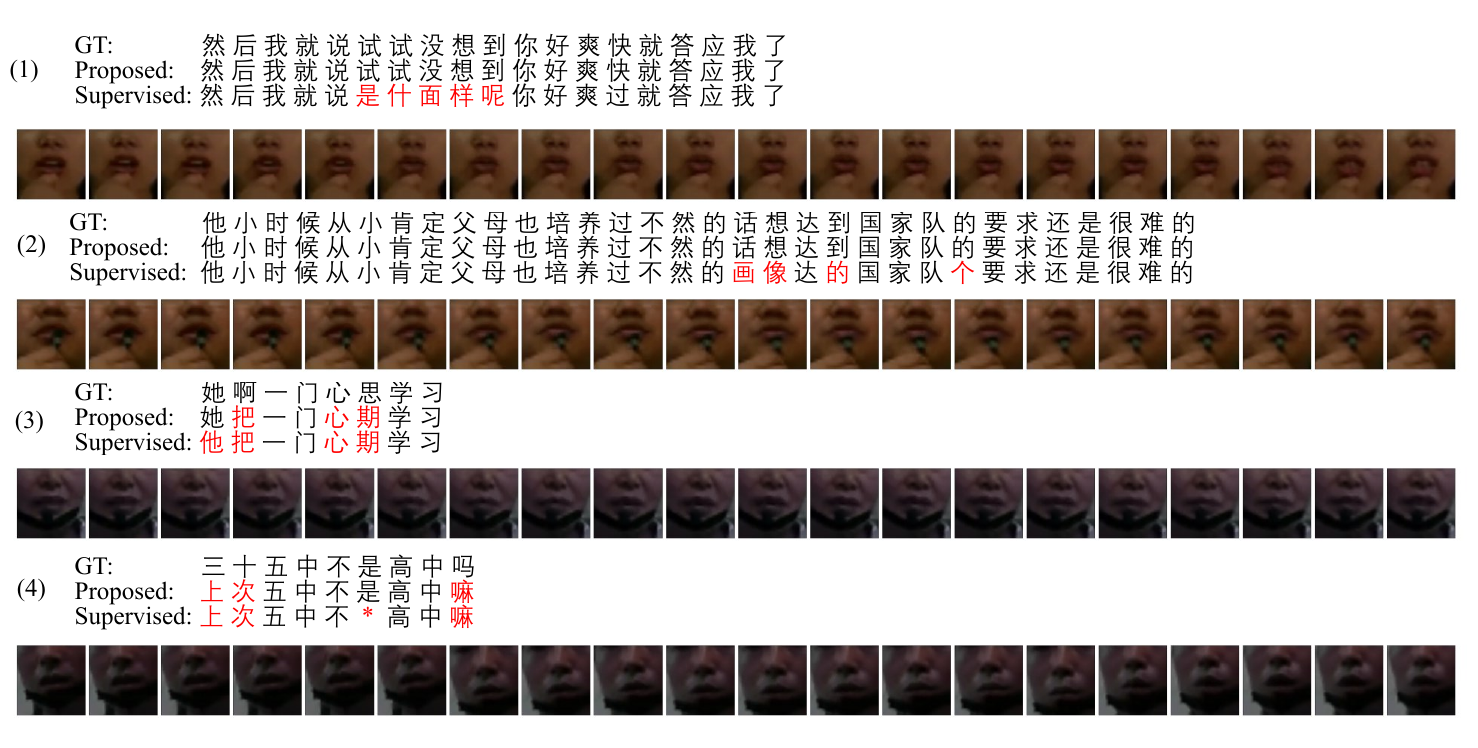}
\caption{Decoding examples of the proposed AV-wav2vec2 model and the end-to-end supervised model (without pre-training), where  GT and red characters denote the ground-truth and wrong results in the output, respectively.}
\label{fig:figure4}
\vspace{-0.2cm}
\end{figure*}

\subsection{Visualization and Analysis}
We randomly select several test sets and decode them using both self-supervised and supervised models, and the visualization is shown in Fig.~\ref{fig:figure4}.
We find that for longer sentences, the use of the self-supervised model is effective in reducing the CER, while for shorter sentences, the CER is relatively high. This is due to the fact that shorter sentences are not beneficial for the model to learn representative contextual information, which can lead to higher error rates. This phenomenon was also found in~\cite{shi2022learning,zhu2022vatlm}.
The MISP2021-AVSR dataset was recorded in a home TV scenario with the presence of multiple speakers, TV sounds, light switches, and other interferences, which pose a challenge to the speech recognition task.
Lip feature extraction is also quite  challenging in case of low brightness.
In addition, for the Chinese dataset, there is also the problem of polyphony, e.g., more than one character having the same pronunciation, which is a common source of the CER. Self-supervised modeling can reduce such errors to some extent, but cannot eliminate them completely.

\section{Conclusion}

In this paper, we first proposed the multi-modal multichannel self-supervised model called AV-wav2vec2, which employs video and multichannel speech inputs.
We found that providing targets for self-supervised pre-training models with speech features from different channels can improve the performance of multi-modal multichannel speech recognition, as both spatial and temporal speech information can be somehow leveraged.
Second, it was shown that utilizing more single-channel speech data is useful for multi-modal multichannel speech recognition provided a small amount of data.	
It is worth mentioning that for the task of Chinese multi-modal speech recognition, there exist a variety of multiple speakers, TV sounds, light switches, and other interferences. Modeling the lip and speech information is still a big challenge in practice.

\section{Acknowledgments}
This work is supported by the National Natural Science Foundation of China (62101523), Hefei Municipal Natural Science Foundation (2022012), USTC Research Funds of the Double First-Class Initiative (YD2100002008).

\bibliography{aaai24}

\end{document}